\documentclass[epj]{svjour}

\newcommand{\ee}[1]{\mathrm{e}^{#1}}

\usepackage{amsmath}
\usepackage{amssymb}
\usepackage{graphics}
\usepackage{epsfig}

\begin{document}

\title{Phase behavior of three-component ionic fluids}

\author{Andr{\'e} G.\ Moreira\thanks{e-mail: amoreira@mpikg-golm.mpg.de}
and Roland R.\ Netz}

\institute{Max-Planck-Institut f{\"u}r Kolloid- und Grenzfl{\"a}chenforschung,
14424 Potsdam, Germany}

\date{Received: date / Revised version: date}

\abstract{
  We study the phase behavior of solutions
  consisting of positive and negative ions of 
  valence $z$ to which a third ionic species of valence $Z>z$ is added.
  Using a discretized Debye-H{\"u}ckel theory, we 
  analyze the phase behavior of such systems 
  for different values of the ratio $\bar{Z} \equiv Z/z$. 
  We find, for $\bar{Z}>1.934$, a
  three-phase coexistence region and, for $\bar{Z}>2$, a closed
  (reentrant) coexistence loop at high temperatures.
  We characterize the behavior of these ternary ionic mixtures
  as function of charge asymmetry and temperature,
  and show the complete phase diagrams for the experimentally relevant
  cases of $\bar{Z}=2$ and $\bar{Z}=3$, corresponding 
  to addition of divalent and trivalent ions to monovalent ionic fluids,
  respectively. 
}

\maketitle

\section{Introduction}

In many areas of Physics, Chemistry, and Biology electrostatic forces play
an important role in determining system properties. This becomes particularly
true when dealing with ionic fluids, i.e., 
fluids consisting of dissociated cations and anions, since in most cases
the Coulomb interaction is the dominant interaction between the particles.
Ionic mixtures have been intensely debated not only because
of their interesting behavior (for recent reviews, see 
\cite{sengers,fisher,pitzer,weingaertner,stell}),
but also because a better understanding of such systems is the 
first step towards a more rigorous theoretical 
treatment of more complicated systems like
charged colloids and polyelectrolytes.

It has been shown experimentally that ionic solutions exhibit
liquid-liquid phase coexistence, terminated by a critical point,
similar to ordinary (nonionic) fluids\cite{kirshenbaum}.
Theoretically, such a demixing transition has been rationalized
in terms of an effective attraction between the ions\cite{mcquarrie,friedman}
which is due to charge screening, first introduced by Debye and H{\"u}ckel\cite{debye}.
For simple inorganic salts, such as \mbox{NaCl}, this
critical point occurs at temperatures above $3000$~K\cite{kirshenbaum}
and thus precludes precise measurements.
More detailed experiments became available with the advent of
large organic ions, which show critical points at temperatures
of $414$~K\cite{pitzer2} and, more recently, at $317$~K\cite{singh}.

The aforementioned examples correspond to binary mixtures of cations
and anions of the same valence. In this article we will
look at three-component ionic mixtures, i.e.,
we will consider solutions of $z \! : \! z$ salts to which a third component
with valence $Z$ is added (keeping total charge neutrality). 
This situation often occurs in real systems, deserving for this reason
a closer look. For instance, in the cellular medium several ionic
species with different valences (like Cl$^-$, Na$^+$ and Ca$^{2+}$) are
simultaneously present. As another example, the presence of charged impurities
can change dramatically the behavior of charge-stabilized
colloidal suspensions\cite{note-tata}. With this in mind, 
adding a third component to an ionic fluid is
the natural extension of previous work and, as we will show, the resulting
phase behavior is quite rich: it is our hope that this will motivate
experiments testing our predictions.

The theoretical tools that are usually employed to treat ionic fluids
are the Debye-H{\"u}ckel theory\cite{debye} (or some modified version of it)
and liquid state theory with some special approximation\cite{jp-hansen}
like the hypernetted chain (HNC) or the mean-sphe\-rical approximation (MSA).
For instance, Caccamo\cite{caccamo}
and Caccamo and Malescio\cite{caccamo-malescio} studied ternary mixtures
with $\bar{Z} \equiv Z/z = 2$, the former using MSA (for low-density mixtures), 
while the latter made a comparison between MSA, HNC and Monte Carlo results
(at high salt densities). Also, three-component mixtures with 
$\bar{Z} = 2$ and 3 have been studied by Kenkare
et al.\cite{kenkare}, where MSA was used to compute the
Coulombic contribution to the free energy. Their phase diagrams,
although presented in a constant-pressure ensemble (in contrast
to our phase diagrams, where we do not constrain the osmotic pressure
to constant values), seem to be generally in accordance with our 
results\cite{nota-cacca,nota-brami}. 

In general, the comparison between computer simulations and 
liquid state theory shows that the latter leads to quite accurate
predictions for the thermodynamic variables of the system. The
same cannot be said about the Debye-H{\"u}ckel theory (cf.\ Table~1):
although it should be exact in the limit of vanishing 
densities, it quickly deviates from the results obtained by computer simulations
as one increases the salt density. However, if one is interested in 
low-density systems, Debye-H{\"u}ckel has the advantage of offering a
very simple free energy that, although not very accurate,
can still be used to study the phase behavior of ionic systems 
without considerable numerical work (in contrast to most approximations
used in liquid state theory).
Besides, Debye-H{\"u}ckel has been shown\cite{netz,note2} to be a Gaussian
theory that can be improved, with the help of field-theoretic methods,
through a \emph{systematic} expansion of the free energy in
cumulants of a fluctuating field (which can be related to the 
electrostatic potential).
For these reasons, we will use here the 
free energy derived from the Debye-H{\"u}ckel theory to describe our
system. 

Our article is presented as follows: in section 2 we briefly 
introduce the theoretical background of our work,
in section 3 we present the resulting phase diagrams for 
various three-component systems and fully characterize their phase
behavior as a function of $\bar{Z}$ and temperature
and finally section 4 
contains some concluding remarks.

\section{Theoretical background}

Let us assume a symmetric ionic system with cations and anions of 
valence $z$ to which a third ionic species of valence $Z$ is
added. Let us also assume that the ions interact only 
via the Coulomb force, or that 
any solvophobic interaction between the ions are weak when compared
to the Coulomb interaction. In this case, following what has been
derived in Ref.~\cite{netz}, the Debye-H{\"u}ckel contribution
to the bulk free energy per volume is
\begin{equation}
  \label{free1}
  f_{\rm DH} =
  -\int \frac{{\rm d} {\bf q}}{(2 \pi)^3}
  \Biggl[ \frac{\kappa^2}{2 \, q^2} -
  \frac{1}{2}\ln \biggl( 1 + \frac{\kappa^2}{q^2} \biggr)
  \Biggr],
\end{equation}
where the electrostatic self-energy is subtracted and
the momentum integral goes over a cube of length $2 \pi/a$.
This accounts for the lattice (with spacing $a$),
implementing in an approximate way
the ionic hard cores. The screening length $\kappa^{-1}$ is 
defined by
\begin{equation}
  \label{kappa}
  \kappa^2 = 4 \pi \ell_B ( z^2 c_+ + z^2 c_- + Z^2 c),
\end{equation}
where the concentrations of the positive and negative ions
of valence $z$ are denoted  by $c_+$ and $c_-$, and the
concentration of the third component (with valence $Z$)
is denoted by $c$. The Bjerrum length 
$\ell_B \equiv e^2 / ( 4 \,\pi \, \varepsilon \, k_B \, T )$ (in SI units)
is the distance at which the electrostatic energy between two elementary
charges equals the thermal energy $k_B \, T$.

Since the integrand in Eq.~(\ref{free1}) is isotropic, we distort the 
integration
volume to a sphere and  obtain after a straightforward integration
\begin{equation}
\label{freeDH}
f_{\rm DH}  = -
\frac{ \kappa^3}{6 \pi^2 } \arctan\left[\frac{\pi }{a \kappa}\right] -
\frac{ \kappa^2}{12  \pi a} +
\frac{ \pi}{12a^3 } \ln \left [1+\frac{a^2 \kappa^2}{\pi^2}\right].
\end{equation}
In the limit $a \rightarrow 0$ this reduces to 
$f _{\rm DH} \simeq - \kappa^3 /12 \pi$, which is
the well-known Debye-H{\"u}ckel limiting law.
The full (Helmholtz) free energy density contains also the ideal entropy
of mixing and reads
\begin{equation}
\label{fullfree}
f = c \ln c+c_+ \ln c_+ +c_- \ln c_- +f_{\rm DH}.
\end{equation}
Global charge neutrality implies that $Z c + z c_+ = z c_-$,
where, without loss of generality,
the third component of valence $Z$ is assumed to be positively charged.
This allows the elimination of one of the concentrations 
from the free energy in Eq.~(\ref{fullfree}): in the following, we choose
to eliminate $c_-$.
For representing our results, we introduce the usual rescaled
temperature $T^* \equiv a/ \ell_B z^2$ and the volume fractions
$\phi^+ \equiv a^3 c_+$ and $\phi \equiv a^3 c$ (notice
that the usual definition of volume fraction---normally represented
as $\eta$---is related to our definition through
$\phi = 6 \, \eta / \pi$).

If one of the three species is absent, the system reduces to 
a two-component system, which has been studied (using the same
free energy) in Ref.~\cite{netz}. 
If  $\phi = 0$, we obtain the lower limiting
critical temperature at $T^*_{\rm LLC} = 0.1776$ with a critical
density of  positive ions of $\phi^+_{\rm LLC}=0.0209$;
if $\phi^+=0$, we obtain the upper limiting critical temperature
at  $T^*_{\rm ULC} = 0.1776 \, \bar{Z}$ and 
$\phi_{\rm ULC} = 0.0418/(1+\bar{Z})$. Clearly, since $\bar{Z}>1$,
we have $T^*_{\rm ULC} > T^*_{\rm LLC}$. These temperatures will
be used as ``guidelines'' in the presentation of the phase behavior of 
three-component systems.

In order to obtain the phase diagrams, we calculate the chemical potentials
$\mu$ and $\mu^+$ from the free energy Eq.~(\ref{fullfree}) by taking
derivatives with respect to $c$ and $c_+$, respectively. We also calculate
the osmotic pressure $p$, which can be easily shown to be
$p = - f + c \, \mu + c_+ \, \mu^+$ (in units of $k_B T$).
We look for instabilities in the free energy through
the sign of the determinant of the (symmetric) matrix
\begin{equation}
  J = 
  \begin{pmatrix}
    \frac{\partial^2 f}{\partial c^2} & 
    \frac{\partial^2 f}{\partial c \partial c_+} \\
    \frac{\partial^2 f}{\partial c_+ \partial c} &
    \frac{\partial^2 f}{\partial c_+^2} & 
  \end{pmatrix}
\end{equation}
in the plane $(c,c_+)$ (or equivalently in the plane
$(\phi,\phi^+)$) at fixed values of $T^*$.
The global convexity of the free energy implies that $\det J$ should be always
positive: when this is not verified, one is in a region where phase separation
occurs. The coexisting phases are then found through
the Maxwell construction\cite{callen}, i.e., one has to find the
$c$ and $c_+$ in each of the coexisting phases
that lead to values of $\mu$, $\mu^+$ and $p$ that are same for all
phases in coexistence. This procedure yields the phase diagrams
that we now present.

\section{Phase diagrams}

\begin{figure}[t]
  \epsfig{file=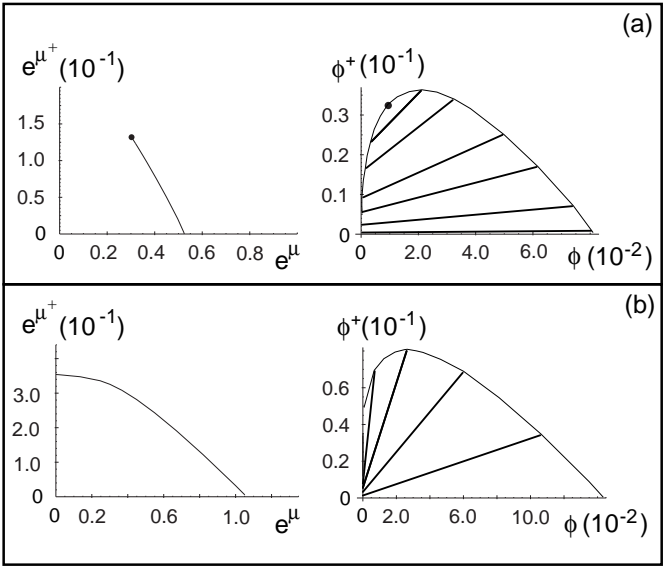,width=88mm}
  \caption{Phase diagrams at constant reduced temperature 
    $T^* \equiv a / \ell_B z^2$
    for a three-component system with a valence 
    ratio $\bar{Z}=1.8$---representative of systems with
    $\bar{Z} =Z/z < 1.934$---as 
    function of $\ee{\mu}$ and $\ee{\mu^+}$
    (the fugacities of $Z$-valent ions and
    of positive $z$-valent ions, respectively) and as function of
    $\phi \equiv a^3 \, c$ and $\phi^+ \equiv a^3 \, c_+$
    (the volume fractions of $Z$-valent ions and
    of positive $z$-valent ions).
    Filled circles denote critical points. 
    Plots: (a) $T^* \equiv a/ \ell_B z^2 = 0.2500$ 
    (representative of the temperature range 
    $T^*_{\rm LLC} < T^* < T^*_{\rm ULC}$); 
    (b) $T^*=0.1754$ (representative for $T^* < T^*_{\rm LLC}$).}
\end{figure}

In Fig.~1 we present the phase behavior for $\bar{Z} = 1.8$, 
representative of small charge ratios $\bar{Z} < 1.934$.
For $T^*> T^*_{\rm ULC} =0.3196$ no phase separation occurs.
In Fig.~1a we show, to the left, the phase diagram in 
fugacities (which are defined as the exponential 
of the chemical potentials) and, to the right, the phase
diagram in
volume fractions for $T^*=0.2500$, representative
of the range $T^*_{\rm LLC} < T^* <T^*_{\rm ULC} $.
There is one critical point at the terminus of a phase coexistence 
line which emanates from the axis defined by $\phi^+ =0$
(or $\ee{\mu^+}=0$).
At $T^* = T^*_{\rm LLC}$ this critical point hits the 
axis $\phi=0$ (or $\ee{\mu}=0$)
at $\phi^+_{\rm LLC}$, and for lower temperatures 
there is single coexistence line connecting the axes
$\phi^+=0$ and $\phi=0$. This is demonstrated
in Fig.~1b, where we show a phase diagram for $T^*=0.1754$. 

\begin{figure}[t]
  \epsfig{file=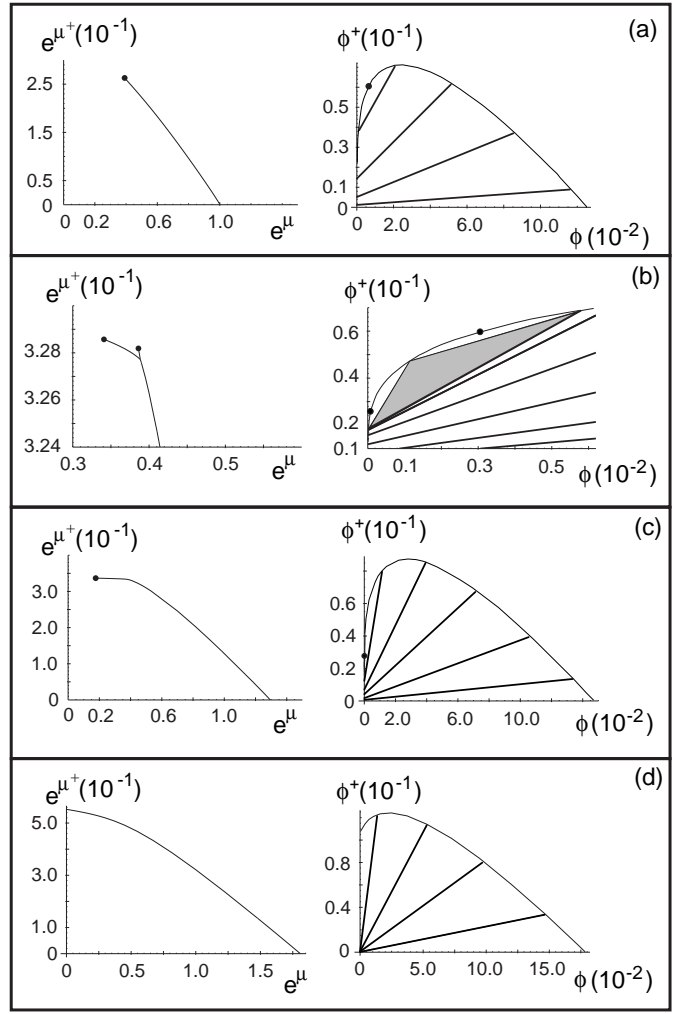,width=88mm}
  \begin{center}
    \caption{Phase diagrams at constant $T^* \equiv a / \ell_B z^2$
      for $\bar{Z} =2$. The notation is the same as in Fig.~1.
      Shaded areas in the volume fraction 
      representation denote three phase coexistence zones. 
      Plots: (a) $T^{*} = 0.2070$ 
      ($T^*_{\rm UCEP} < T^* < T^*_{\rm ULC}$);
      (b) $T^{*} = 0.1800$ ($T^*_{\rm LCEP} < T^* < T^*_{\rm UCEP}$)
      (only shown close to the three phase coexistence zone);
      (c) $T^{*} = 0.1786$ ($T^*_{\rm LLC} < T^* < T^*_{\rm LCEP}$);
      (d) $T^{*} = 0.1429$ ($T^* < T^*_{\rm LLC}$).}
  \end{center}
\end{figure}

The behavior is dramatically altered for larger 
charge asymmetry, viz.\ $\bar{Z} > 1.934$. One important
difference in the phase topology is that such systems always have 
two critical end-points, one at temperature $T^*_{\rm UCEP}$
(upper critical end-point) and another one at $T^*_{\rm LCEP}$
(lower critical end-point). In the temperature range between these
two critical end-points the systems exhibit three phase coexistence.

In Fig.~2 we present
the phase behavior for $\bar{Z}=2$ corresponding, for instance,
to a mixture of a divalent cations with monovalent co- and counter-ions.
For high temperatures, $T^*> T^*_{\rm ULC} = 0.3552$ no phase
separation occurs.
For a temperature range $T^*_{\rm UCEP} <T^*< T^*_{\rm ULC}$ 
there is a single
critical point, as demonstrated  in Fig.~2a for $T^*=0.2070$; 
the phase diagram
is similar to the one shown in Fig.~1a. The upper critical 
end-point is located
at $T^*_{\rm UCEP} = 0.1808$, below which a second critical 
point exists. The resulting phase behavior for
$T^*_{\rm LCEP} <T^*< T^*_{\rm UCEP}$ is similar to
Fig.~2b, which depicts the phase diagram for $T^*=0.1800$,
and where we see two phase coexistence lines, each
terminating at a critical point and merging at a triple
point. The region around this triple point is highlighted 
at the phase diagram in volume fractions (Fig.~2b, right).
At $T^*=T^*_{\rm LCEP}=0.1796$ one of the critical lines
present for $T^*_{\rm LCEP} <T^*< T^*_{\rm UCEP}$ terminates
at the lower critical end-point.
For $T^*_{\rm LLC} < T^* < T^*_{\rm LCEP}$ we are thus left again with 
a single critical point; a representative phase diagram 
is shown in Fig.~2c for $T^* = 0.1786$. Finally, for 
$T^*< T^*_{\rm LLC} = 0.1776$, no critical point is observed;
Fig.~2d shows such a phase diagram for $T^*=0.1429$,
which is similar to the one in Fig.~1b.

\begin{figure}[t]
  \epsfig{file=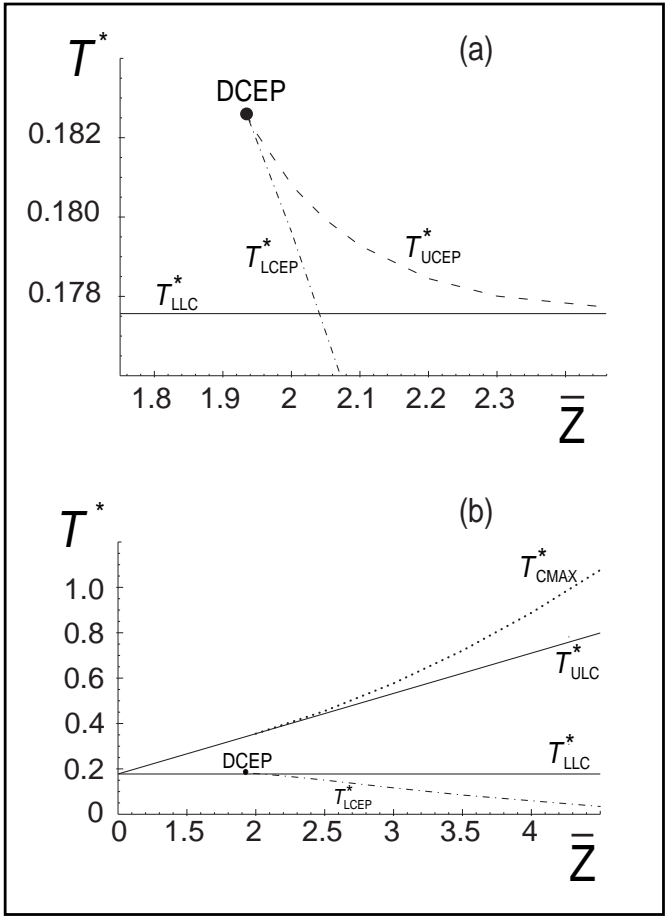,width=88mm}
  \begin{center}
    \caption{(a) Plot of the upper critical end-point temperature 
      $T^*_{\rm UCEP}$ (broken line), the lower critical 
      end-point temperature
      $T^*_{\rm LCEP}$ (dotted-broken line), and the lower 
      limiting critical
      temperature $T^*_{\rm LLC}$ (solid line) as a 
      function of the valence
      ratio $\bar{Z}$ in the vicinity of the double critical 
      end-point (DCEP). 
      (b) Plot of the lower and upper limiting critical temperature, the
      maximal critical temperature ($T^*_{\rm CMAX}$) and the 
      lower critical end-point 
      temperature as a function of $\bar{Z}$; notice the widening 
      of the three phase
      zone and the growth of $T^*_{\rm CMAX}$ as $\bar{Z}$ increases.}
  \end{center}
\end{figure}

The complete phase behavior is summarized in Fig.~3, where we plot the 
various critical temperatures as a function of the valence ratio 
$\bar{Z}$.
In Fig.~3a we show, in the vicinity of $\bar{Z}=2$,
the upper critical end-point temperature $T^*_{\rm UCEP}$
(broken line), the lower critical end-point temperature $T^*_{\rm LCEP}$
(dotted-broken line), and the lower limiting critical temperature
$T^*_{\rm LLC}$ (solid line). The two
lines of critical end-points meet at a double critical end-point at
$\bar{Z}=1.934$,  below which
no triple point occurs. 
On a larger scale (Fig.~3b) it becomes clear that the triple
phase temperature range widens as $\bar{Z}$ increases (note that
$T^*_{\rm UCEP}$ and the double critical end-point cannot be 
distinguished from
$T^*_{\rm LLC}$ at this scale). We also plot
the maximal temperature at which a critical point occurs, 
$T^*_{\rm CMAX}$,
which for $\bar{Z} > 2$ satisfies $T^*_{\rm CMAX} > T^*_{\rm ULC}$. This
leads to closed coexistence loops, as is demonstrated in the following.

In Fig.~4 we show the phase behavior for $\bar{Z}=3$. 
In Fig.~4a we plot the phase diagram for $T^*=0.5747$, which is 
below the maximal critical temperature $T^*_{\rm CMAX} = 0.5771$ 
and above the
upper limiting critical temperature $T^*_{\rm ULC}= 0.5328$. Notice the 
presence of two critical points bounding a closed coexistence loop. 
For $T^*_{\rm UCEP} < T^* < T^*_{\rm ULC}$ we find a phase 
topology with a single
critical point, as depicted in Fig.~4b for $T^*=0.5155$.
In this system the upper critical end-point occurs at a temperature
very close to the lower limit critical temperature 
(this is already clear in
Fig.~3); more precisely, $T^*_{\rm LLC}=0.1775656$ and 
$T^*_{\rm UCEP}=T^*_{\rm LLC} + 1.0 \times 10^{-6}$.
In Fig.~4c, where $T^* = T^*_{\rm LLC} + 0.7 \times 10^{-6}$
($T^*_{\rm LLC} < T^* < T^*_{\rm UCEP}$), we observe
a triple point and two critical points.
Since the lower critical end-point occurs at a lower
temperature than the lower limiting critical point
(cf.\ Fig.~3), we obtain in the range 
$T^*_{\rm LCEP} < T^* < T^*_{\rm LLC}$
one critical point and a triple point, as shown in Fig.~4d for 
$T^* = 0.1667$. Finally, for $ T^*< T^*_{\rm LCEP} = 0.1163$, all
critical points have disappeared and a coexistence line
connects the two phase diagram axes, as shown in Fig.~4e for 
$T^*=0.1000$.

\section{Conclusion}

For small valence ratios ($\bar{Z} \equiv Z/z < 1.934$) we have shown
that the the two critical points
obtained for the limiting cases (i.e., one of the two
positively charged species is absent) are joined by
a continuous critical line. For larger valence ratios
($\bar{Z} > 1.934$) we have obtained more complicated phase diagrams,
with a line of triple points bounded by critical end-points
and closed coexistence rings at $T > T^*_{\rm ULC}$ for
$\bar{Z} > 2$. We have also made a global analysis of three-component
ionic mixtures, where their phase behavior is characterized as a 
function of charge asymmetry (up to $\bar{Z} \sim 5$) 
and temperature.

Denoting by $T^*_{\rm LLC}$ the lower limiting critical temperature
of the symmetric ionic system, we have shown that this
triple line extends, for $\bar{Z} =2$, over a temperature window of
$1.012 \, T_{\rm LLC}^*<T^*< 1.018 \, T^*_{\rm LLC}$
and, for $\bar{Z} =3$, over
$0.66 \, T^*_{\rm LLC} < T^* \lesssim T^*_{\rm LLC}$.
This means that for a $1 \! : \! 1$ organic salt with a critical point at
$T \simeq 400$~K this temperature window would have a $2.4$~K width when
adding divalent ions and a $136$~K width when adding trivalent ions.
The triple line should in principle be observable with a suitable choice of
mono-, di- and trivalent organic ions.
However, one should be careful when examining the numbers presented here.
Although the Debye-H{\"u}ckel theory successfully predicts for ionic systems
the existence of a liquid-liquid critical point and the high asymmetry
of their phase diagrams, it has some limitations, as we now discuss.

\begin{figure}[t]
  \epsfig{file=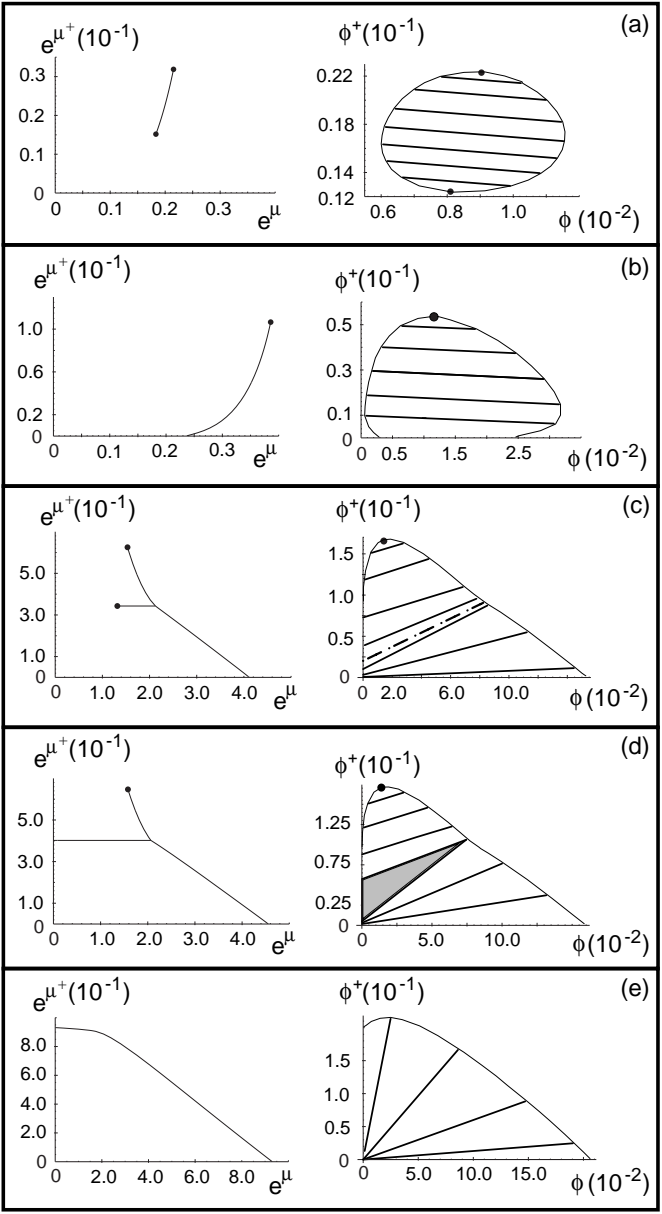,width=88mm}
  \begin{center}
    \caption{Phase diagrams at constant $T^* \equiv a / \ell_B z^2$
      for $\bar{Z}=3$. The notation is the same as in Figs.~1 and 2.
      Plots:
      (a) $T^{*} = 0.5747$ ($T^*_{\rm ULC} < T^* < T^*_{\rm CMAX}$);
      (b) $T^{*} = 0.5155$ ($T^*_{\rm UCEP} < T^* < T^*_{\rm ULC}$); 
      (c) $T^{*} = 0.1775663$ ($T^*_{\rm LLC} < T^* < T^*_{\rm UCEP}$)
      (in volume fraction the three phase coexistence is 
      represented by a dashed line);
      (d) $T^{*} = 0.1667$ ($T^*_{\rm LCEP} < T^* < T^*_{\rm LLC}$); 
      (e) $T^{*} = 0.1000$ ($T^* < T^*_{\rm LCEP}$).}
  \end{center}
\end{figure}

One problem is the fact that the theory 
is quite sensitive to the way that the
hard cores are treated: we chose to include them through a lattice
of spacing $a$. Alternatively, one could add to the free energy
Eq.~(\ref{fullfree}) an extra term to account for the hard-core repulsion 
(for instance a free-volume approximation term)
and get the electrostatic contribution $f_{\rm DH}$ either by
integrating Eq.~(\ref{free1}) without the cut-off (which results immediately
in the Debye-H{\"u}ckel limiting law) or by doing a ``charging process''
after solving of the linearized Poisson-Boltzmann equation with the
appropriate boundary conditions for the hard 
cores\cite{debye,fisher-levin,size}. 
These methods yield similar phase diagrams, but with different values of
temperature and density for the critical point. The use of a lattice
is the most convenient for the field-theoretic treatment, although it is
certainly not the best way\cite{tcp-ocp}.
This, as well as some simplifications done 
during the calculations (like performing the integral 
in Eq.~(\ref{free1}) in the sphere instead of the initial cube 
of size $2 \pi /a$) are expected to affect the final results.

The direct comparison of the internal energy per particle that follows
from Eq.~(\ref{fullfree}), viz.
\begin{equation}
  \label{internal}
  u \equiv \frac{U_{\mathrm{ex}}}{N \, k_B \, T} = 
  -\frac{a^3 \kappa^3}{8 \, \pi \, \phi^+} \arctan 
  \Bigl(\frac{\pi}{a \, \kappa} \Bigr),
\end{equation}
with results from other approaches is shown in Table~1 
(in Eq.~(\ref{internal}), $\kappa$ is the inverse screening length defined in 
Eq.~(\ref{kappa}), $a$ is the lattice spacing, corresponding to the
ionic diameter, and $\phi^+$ is the volume fraction 
of the positive $z=1$ ions). In the first part of this
table (``binary''), we compare the values of $u$ from Monte Carlo 
simulations, HNC, MSA\cite{rasaiah-card-valleau} and
Debye-H{\"u}ckel limiting law (DH) for a simple $1 \! : \! 1$ ionic mixture
(where $\phi=0$, i.e., the $Z$-valent ions are not present).
Although our results are better than the pure DH results, it is clear that
our theory needs improvement at higher volume fractions. In the second part
of table~1 (``ternary''), the same comparison is done for a three-component
mixture with $\bar{Z}=2$ with results obtained by Caccamo\cite{caccamo}
(notice that the
value of $\phi^+$ is in the region where our phase diagrams are shown).
This shows that our theory underestimates the value
of the internal energy of the system, which
indicates that, beyond the problem with the hard cores, 
a more refined free energy is needed in order to capture correctly all features
of the ionic fluids\cite{pairs}. This idea is reinforced by the
discrepancy between the critical parameters for a $1 \! : \! 1$
salt obtained through
MC simulations\cite{simul} (which lie in the
range $T^*_{\rm C} = 0.049$--$0.070$ and
$\phi^{+}_{\rm C}= 0.013$--$0.040$) with the ones obtained with the
free energy used here ($T^*_{\rm C} = 0.178$ and $\phi^{+}_{\rm C} = 0.021$).

\begin{table}[t]
  \begin{tabular}{cccccc}
    \hline
    $\phi^+$ & MC & HNC & MSA & DH & Eq.~(\ref{internal}) \\ \hline
    binary & & & & \\
    \mbox{$4.2 \! \times \! 10^{-4}$} & $0.1029$ & $0.1014$ & $0.0992$ & 0.1120 & $0.1090$ \\
    \mbox{$4.8 \! \times \! 10^{-3}$} & $0.2739$ & $0.2714$ & $0.2675$ & 0.3780 & $0.3438$ \\
    \mbox{$1.9 \! \times \! 10^{-2}$} & $0.4341$ & $0.4295$ & $0.4264$ & 0.7651 & $0.6276$ \\
    \mbox{$4.6 \! \times \! 10^{-2}$} & $0.5516$ & $0.5447$ & $0.5405$ & 1.1736 & $0.8610$ \\
    \hline
    ternary & & & & \\
    \mbox{$2.1 \! \times \! 10^{-3}$} & --- & $1.520$ & $1.301$ & 2.327 & $1.913$ \\
    \hline
  \end{tabular}
  \caption{Comparison between (minus) the excess internal energy per particle
    from Monte Carlo simulations (MC), hypernetted chain (HNC), 
    mean spherical approximation (MSA), Debye-H{\"u}ckel limiting law (DH) 
    and Eq.~(\ref{internal}) ($\phi^+ = a^3 \, c_+$, as defined in Section~2).
    In the first part of the table (``binary'') the comparison
    is done for a simple $1 \! : \! 1$ electrolyte with $T^* = 0.59$,
    where the data for MC, HNC and MSA are taken from Table~II of
    Ref.~\cite{rasaiah-card-valleau}.
    In the second part of the table (``ternary''), the same comparison is done for a
    three component mixture with $\bar{Z}=2$, $T^*=0.294$ and $\phi = \phi^+$; 
    the data for HNC and MSA
    are taken from Table~I of Ref.~\cite{caccamo}.}
\end{table}

With this in mind, one should regard the results we obtained
as being of ``zeroth-order.'' As previously mentioned, the free
energy used here is the lowest-order result following a
systematic expansion. Changes in the phase behavior are expected
as one takes into account higher order terms in the expansion,
but as shown in Ref.~\cite{netz} for $Z \! : \! z$ mixtures, the inclusion
of terms up to the eighth cumulant in the fluctuating field 
affects the phase diagrams appreciably only for
values of $\bar{Z} \equiv Z / z \gtrsim 5$.
For this reason, and in spite
of the limitations of the Debye-H{\"u}ckel theory, we assume that
the topology of the phase diagrams shown here are correct.

\begin{acknowledgement}
We thank P.\ van der Schoot for stimulating discussions
and H.\ Weing{\"a}rtner and J.\ M.\ H.\ Levelt Sengers for
useful correspondence.
AGM acknowledges the support from the Portuguese FCT 
through grant PRAXIS XXI/BD/13347/97.
\end{acknowledgement}

\end{document}